\documentclass{pasj00}
\draft

\begin{document}
\SetRunningHead{T. Ohtani, S. S. Kimura, and T. Tsuribe}{Role for the Inner Disks in 
Mass Acceretion Rate to the Star}

\title{The Role for the Inner Disk in Mass Accretion to the Star in the Early Phase of Star Formation}

\author{Takuya \textsc{Ohtani}\altaffilmark{1}, Shigeo S. \textsc{Kimura}\altaffilmark{1}, Toru \textsc{Tsuribe}\altaffilmark{1}} 
\altaffiltext{1}{Department of Earth and Space Science, Graduate School of Science, Osaka University, 1-1 Machikaneyama-cho, Toyonaka, Osaka 560-0043}
\email{ohtani@vega.ess.sci.osaka-u.ac.jp, kimura@vega.ess.sci.osaka-u.ac.jp, tsuribe@vega.ess.sci.osaka-u.ac.jp}
\and 
\author{Eduard I. \textsc{Vorobyov}\altaffilmark{2,3}
}
\altaffiltext{2}{Department of Astrophysics, University of Vienna, 1170, Austria} 
\altaffiltext{3}{Research Institute of Physics, Southern Federal University, Rostov-on-Don, 344090, Russia}
\email{eduard.vorobiev@univie.ac.at}

\maketitle

\begin{abstract}
A physical mechanism that drives FU Orionis-type outbursts is reconsidered. 
We study the effect of inner part of a circumstellar disk covering a region from near the central star to the radius of approximately $5$ AU (hereafter, the inner disk). 
Using the fluctuated mass accretion rate onto the inner disk $\dot{M}_{\rm out}$, we consider the viscous evolution of the inner disk and the time variability of the mass accretion rate onto the central star $\dot{M}_{\rm in}$ by means of numerical calculation of an unsteady viscous accretion disk in a one-dimensional axisymmetric model. 
First, we calculate the evolution of the inner disk assuming an oscillating $\dot{M}_{\rm out}$. 
It is shown that the time variability of $\dot{M}_{\rm in}$ does not coincide with $\dot{M}_{\rm out}$ due to viscous diffusion. 
Second, we investigate the properties of spontaneous outbursts with temporally constant $\dot{M}_{\rm out}$. 
Outburst occur only in a limited range of mass accretion rates onto the inner disk $10^{-10}<\dot{M}_{\rm out}< 3\times 10^{-6}~{\rm M}_{\odot} {\rm yr}^{-1}$ due to gravo-magneto limit cycle (GML). 
Finally, we discuss the case with a combination of episodic $\dot{M}_{\rm out}$ and accretion outbursts cause by the GML in the inner disk. 
The GML can drive accretion outbursts onto the star even for the case of fluctuating $\dot{M}_{\rm out}$, although fluctuations of $\dot{M}$ decay during transmitting the inner disk inwards. 
We newly identified two modes of outburst 
which are spontaneous one and stimulated one. 
In a stimulated mode of outburst, $\dot{M}_{\rm out}$ does appear directly in $\dot{M}_{\rm in}$ (the latter defining the stellar accretion luminosity). 
In a spontaneous mode of outburst, $\dot{M}_{\rm out}$ appears as the interval between outbursts. 
\end{abstract}

\section{Introduction}
In the early phase of star formation, it is thought that mass is accreted by the central star. 
FU Ori-type objects provide direct observational evidence for the mass outbursts with short duration time ($\sim100~{\rm yr}$). 
A number of theoretical models have been proposed to explain FU Ori-type outbursts. 
There are famous three types of models, including the thermal instability (TI) model (Bell \& Lin \yearcite{BellLin1994}), gravitational instability (GI) model (Vorobyov \& Basu \yearcite{VoroBasu05}, \yearcite{VB06}), and gravo-magneto limit cycle (GML) model (Armitage et al. \yearcite{Armitage01}, Zhu et al. \yearcite{Zhu09}, and Martin \& Lubow \yearcite{ML11}).  

Bell \& Lin (\yearcite{BellLin1994}) discussed whether thermal instability in the inner region of the disk can explain outbursts. 
They obtained the steady state solutions of the accretion disk, and they showed the results on $\dot{M}$ versus $\Sigma$ plane. 
In their model, they succeeded in explaining time scales of outburst with taking into account of vertical convection of the disk and radial heat flux. 
They predict that outbursts occur in the region $r<0.1~{\rm AU}$ and $T\sim 10^{4}~{\rm K}$. 
However, the temperature which TI occur is too high for protostellar disk if we consider the early phase of disk formation. 
Zhu et al. (\yearcite{Zhu07}, \yearcite{Zhu08}) pointed out that TI model cannot work for FU Ori because the region with high accretion rate is required to extend to $0.5~{\rm AU}$. 

Using two-dimensional simulations of star and disk formation, 
Vorobyov \& Basu (\yearcite{VoroBasu05}, \yearcite{VB06}) discussed the GI model. 
It is shown that formed via disk gravitational fragmentation at $\sim 100 \rm AU$ in the early phase of the disk formation can be driven into the central region and cause the mass accretion outbursts. 
However, due to the use of a sink cell, Vorobyov \& Basu (\yearcite{VoroBasu05}, \yearcite{VB06}) could not follow the disk evolution in the inner $5$ AU.
Although they assume the mass accretion rate onto their sink cell as that onto the star, the validity of this assumption is unclear. 
To help resolve this problem, in this paper we consider the inner region of the disk (hereafter the inner disk). 

In the third GML model of outbursts in the inner region of the disk, magneto-rotational instability (MRI) is used to explain outbursts (Armitage et al. \yearcite{Armitage01}, Zhu et al. \yearcite{Zhu09}, \yearcite{Zhu10a}, \yearcite{Zhu10b}, and Martin \& Lubow \yearcite{ML11}). 
In GML model, the disk has two states; which are 
MRI-active state when $T>T_{\rm crit}$ and MRI-inactive state when $T<T_{\rm crit}$\footnote{MRI is active when gas are well coupled with the magnetic field with high ionization degree when temperature is high ($T>T_{\rm crit}$), where most of dust components evaporate, thus eliminating a major source of sink for current-carring electrons. }. 
In GML model, spontaneous outbursts are driven by the state transition between MRI active and inactive state. 
The transition is caused by mismatches of mass accretion rate between inner and outer region of the disk. 
In the view point of angular momentum transport, MRI is dominant in the inner region ($\sim 0.2~{\rm AU}$) and GI is dominant in the outer region ($\sim1~{\rm AU}$) of the disk. 
GML model of outbursts is demonstrated by numerical calculation, 
where outburst is associating with the time variation of MRI active/inactive state in the inner region of the disk (Armitage et al. \yearcite{Armitage01}, Zhu et al. \yearcite{Zhu09}, \yearcite{Zhu10a}, and \yearcite{Zhu10b}). 

Previous studies of the GML model were limited to the case of a temporally constant mass accretion rate onto the inner disk.
According to studies of the GI model, accretion rate onto the inner disk is expected to be varying in time. 
In this paper, we consider the role for the inner disk in time variation of mass accretion rate onto the inner disk. 
The viscous evolution of the inner disk ($0.2\sim 5.0~{\rm AU}$) is solved with taking into account both the outbursts driven by GML and the time variable mass accretion rate onto the inner disk. 
In order to understand the time variability of mass accretion rate onto the star, 
first we consider the above two processes independently. 
In section 2, using a simple model of viscous accretion disk, we show the accretion rate at near the surface of protostar as the responses to the episodic accretion at the outer boundary. 
In section 3, features of outbursts driven by GML are shown with temporally constant accretion rate onto outer boundary of the inner disk. 
In section 4, we obtain the time variability of accretion rate as a result of combination of outer episodic accretion and GML.  
We discuss implications of our results and future work in section 5.  
Our results are summarized in section 6. 

\section{Response to the Oscillating Accretion onto the Inner Disk}
\subsection{Assumptions and Basic Equations}
The accretion disk model that we use in this paper 
describes on the unsteady, viscous accretion process of the circumstellar disk by assuming that the disk is axisymmetric and geometrically thin. 
All the equations are integrated in the vertical direction in cylindrical coordinates ($r, \phi, z$). 
The system is described by a one-dimensional equation dependent on time $t$ and radial distance $r$. 
We apply this disk model to the inner disk. 
First, we investigate the simple response of the inner disk to the oscillating accretion onto the outer boundary of the inner disk. 
Note that mass accretion rate onto the outer region in our model corresponds to the accretion rate onto the sink cell in previous works on the GI model (e.g., Vorobyov \& Basu \yearcite{VB10}). 

We assume that the pressure gradient force is negligible, 
and the accretion speed within the disk is sufficiently slow as 
\begin{equation}
|\frac{c_{\rm s}^2}{\Sigma}\frac{\partial \Sigma}{\partial r}/{\frac{GM_*}{r^2}}|<< 1~{\rm and}~|v_{\rm r}|<<v_{\phi},
\label{rotation}
\end{equation}
where $\Sigma$ is the surface density, $v_{\rm r}$ is the radial velocity, and $v_{\rm \phi}$ is azimuthal velocity which is assumed to be Keplerian  $v_{\phi}=\sqrt{{GM_*}/{r}}$.  
As a result, viscous evolution of the surface density $\Sigma$ is described by 
\begin{equation}
\frac{\partial \Sigma}{\partial t}-\frac{1}{r}\frac{\partial}{\partial r}\left(\frac{\dot{M}(r)}{2\pi}\right)=0,\quad{\rm with}\quad\dot{M}(r)=-\frac{2\pi}{\frac{\partial(rv_{\phi})}{\partial r}}\frac{\partial}{\partial r}(\nu \Sigma r^3\frac{\partial \Omega}{\partial r}), 
\label{diffusiveeq}
\end{equation}
where $\Omega=v_{\phi}/r$ is the angular velocity and $\nu$ is viscous coefficient (e.g., Pringle \yearcite{Pringle} and \S 7.2 in Hartmann \yearcite{Hartmann}). 
A hydrostatic equilibrium in the $z$-direction is assumed. 
Then the scale height $h$ is given by $h=c_{\rm s}/\Omega$. 
As angular momentum transport, the standard $\alpha$ prescription for viscous disk (\cite{Pringle}, \cite{Shakura}) is employed. 
By this prescription, viscous coefficient $\nu$ in equation (\ref{diffusiveeq}) is given by $\nu =\alpha c_{\rm s}h$. 
In this section, we use the simplified treatmeant on $\alpha$ and $c_{\rm s}$ 
with temporally and spatially constant $\alpha=0.1$, and temporally constant $c_{\rm s}=0.63 (r/1{\rm AU})^{-1/4}~[{\rm km}/{\rm s}]$. 

We solve equation (\ref {diffusiveeq}) numerically for time evolution of $\Sigma$ and $\dot{M}(r)$ in $r_{\rm in}<r<r_{\rm out}$, where $r_{\rm in}=0.2~{\rm AU}$ and $r_{\rm out}=5~{\rm AU}$. 
Mass of the central star $M_*$ is set to be $0.5~{\rm M}_{\odot}$. 
The diffusion time $t_{\rm diff}(r)=r^2/{\nu(r)}$ at $r_{\rm out}=5~{\rm AU}$ is about $17,500~\rm yr$ in this case. 
For the mass accretion rate onto the disk outer edge, supposedly mimicking a periodic effect of GI in the outer disk, we adopt the following form 
\begin{equation}
\dot{M}_{\rm out}(t)=\dot{M}_{0}/{[{\rm cos}(\frac{2\pi}{t_{\rm osci}}\times t)+(1+\delta)]},
\label{Mdotouter}
\end{equation}
where $\dot{M}_{0}$, $t_{\rm osci}$, and $\delta$ are constant parameters. 
Using equation (\ref{Mdotouter}), time averaged value of mass accretion rate at the outer boundary in one period of oscillation is given as $<\dot{M}(r_{\rm out})>_{\rm osci}\equiv \int^{t_0+t_{\rm osci}}_{t_0}\dot{M}_{\rm out}dt \sim \dot{M}_{0}/\sqrt{2\delta}$. 
Equation (\ref{Mdotouter}) includes the case of constant $\dot{M}_{\rm out}=M_{0}/2$ with $t_{\rm osci}=\infty$ and $\delta<<1$. 
Equation (\ref{Mdotouter}) with $\dot{M}_{0}=5\times 10^{-8}~{\rm M}_{\odot} {\rm yr}^{-1}$ and $\delta=0.01$ is used in this section. 
As the inner boundary condition, we use a free  boundary, ${\partial \Sigma}/{\partial r}=0$ at $r=r_{\rm in}$. 
We set the grid number $N_r=37$, $\Delta r$ equally divided in logarithmic space, 
and $dt=C_{\rm safe}{\Delta t}_{\rm diff}=C_{\rm safe}(\Delta r)^2/\nu$, where $C_{\rm safe}=10^{-2}$. 
For the initial conditions, we set power law distributions for $\Sigma=\Sigma_0 r^{-1}$. 
However, initial conditions are unimportant because the inner disk forgets them after several diffusion time scales. 
\subsection{Result}
\begin{figure}
\begin{center}
\FigureFile(80mm,80mm){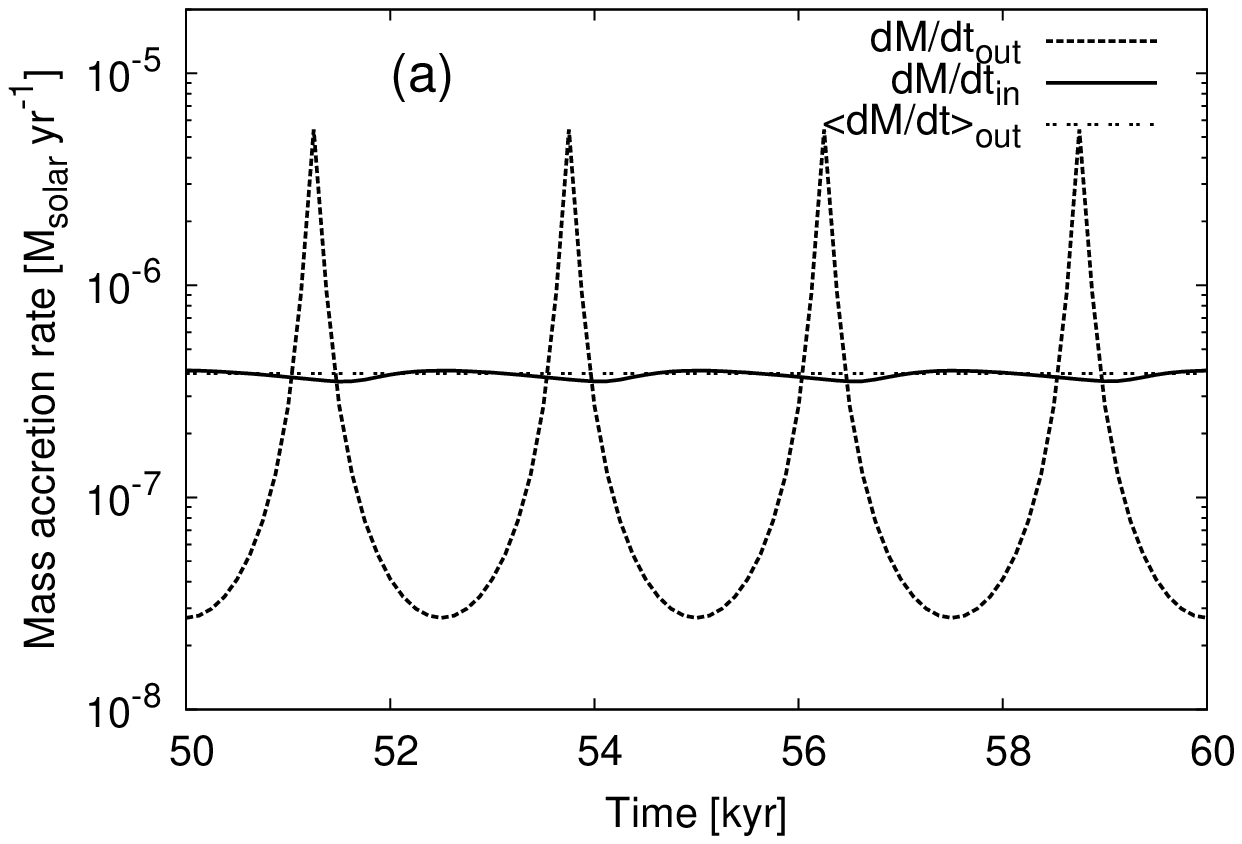}
\FigureFile(80mm,80mm){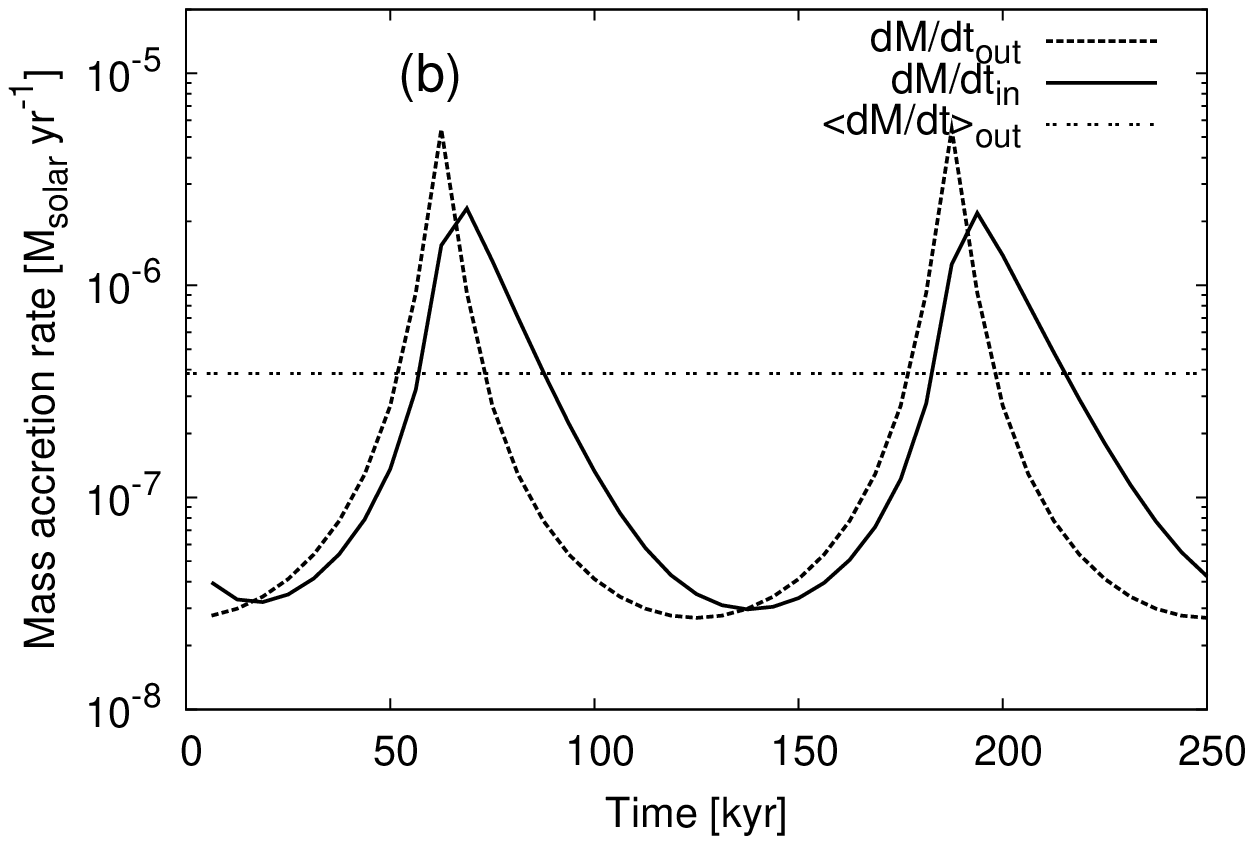}
\end{center}
\caption{Simple responses of the mass accretion rate at $r=r_{\rm in}$ to the oscillated mass accretion at the outer boundary for the cases with $t_{\rm osci}=2,500~{\rm yr}$ (a) and $125,000~{\rm yr}$ (b). Constant $\alpha =0.1$ and temporally constant $c_{\rm s}=0.63 (r/1{\rm AU})^{-1/4}~ [{\rm km}/{\rm s}]$ are used. For these parameters, $t_{\rm diff}(r_{\rm out})=17,500~{\rm yr}$.}
\end{figure} 
The time evolution of the mass accretion rate at $r_{\rm out}$ and $r_{\rm in}$ are shown in Figure 1 for the case with $t_{\rm osci}=2,500~{\rm yr}$ (a) and $t_{\rm osci}=125,000~{\rm yr}$ (b). 
From Figure 1 (a), where $t_{\rm osci}<t_{\rm diff}(r_{\rm out})$, it is seen that mass accretion rate at $r_{\rm out}$, $(dM/dt)_{\rm out}$ (dotted line) has a large oscillation amplitude while mass accretion rate at $r_{\rm in}$, $(dM/dt)_{\rm in}$ (solid line) has a small amplitude. 
This means that the inner disk acts to smooth out oscillations in the mass accretion onto the disk's outer edge and the true accretion rate onto the star exhibits little time variability. 
From Figure 1 (a), it is also seen that mass accretion rate at $r_{\rm in}$ is almost the same as the time-averaged value of mass accretion rate at the outer boundary $<\dot{M}(r_{\rm out})>_{\rm osci}\sim \dot{M}_{0}/\sqrt{2\delta}$, rather than $\dot{M}(r_{\rm out})$. 
This means that if $t_{\rm osci}<t_{\rm diff}(r_{\rm out})$, $\dot{M}(r)$ approaches $<\dot{M}(r_{\rm out})>_{\rm osci}$ in small $r$ limit. 
In Figure 1 (b), where $t_{\rm diff} (r_{\rm out})<t_{\rm osci}$, it is seen that the oscillation amplitude of mass accretion rate is similar at the inner and outer boundaries of the inner disk, 
although the oscillation of mass accretion rate at the inner boundary slightly decays due to viscous diffusion in the inner disk. In summary, mass accretion rate onto the protostar tends to be different from that onto the sink cell due to viscous diffusion. 
Mass accretion rate onto the protostar approaches the time-averaged value $<\dot{M}_{\rm out}>_{\rm osci}$ when $t_{\rm osci}<t_{\rm diff}$ and the time variability of the mass accretion rate is weakly affected when $t_{\rm diff}<t_{\rm osci}$. 

\section{Spontaneous Outburst with Steady $\dot{M}_{\rm out}$}
\subsection{Formulation}
In this section, we investigate features of outbursts driven by the GML model. 
Time variability of mass accretion rate at $r_{\rm in}$ is investigated using steady mass accretion rate onto the outer boundary $r_{\rm out}$ of the inner disk $\dot{M}_{\rm out}$ by solving equation (\ref{diffusiveeq}). 
In this subsection, the effects of MRI and GI are effectively included in $\alpha$ as 
\begin{equation}
\alpha=\alpha _{\rm M}+\alpha _{\rm G}, 
\label{alpha}
\end{equation}
where $\alpha_{\rm M}$ is a viscous parameter which mimics angular momentum transport induced by the MRI, and $\alpha_{\rm G}$ is viscous parameter which mimics angular momentum transport induced by gravitational instabilities. 
As for $\alpha_{\rm M}$, we take into account for the MRI active/inactive state by $\alpha_{\rm M,on}$ and $\alpha_{\rm M, off}$ as,  
\begin{equation}
\alpha_{\rm M}=
\left\{
\begin{array}{c}
\alpha_{\rm M,on}\equiv 10^{-2} \quad (\rm if \quad T>T_{\rm crit})\\
\alpha_{\rm M,off}\equiv 10^{-5} \quad (\rm if \quad T<T_{\rm crit}),
\label{MRIalpha}
\end{array}
\right.
\end{equation}
where $T_{\rm crit}=1400~{\rm K}$ $(c_{\rm s,crit}=2.6~{\rm km}/{\rm s})$ is the critical temperature (see also Zhu et al. \yearcite {Zhu09}).  
A small but nonzero value of $\alpha_{\rm M,off}$ mimics a finite amount of angular momentum transport in the MRI active surface layer around the dead zone in MRI inactive region. 
Since the exact value of $\alpha_{\rm M,off}$ is not well known, we assume this value, which is sufficiently smaller than $\alpha_{\rm M,on}$. 
Viscous parameter $\alpha_{\rm G}$ is assumed to be 
\begin{equation}
\alpha _{\rm G} =
\left\{
\begin{array}{c}
\eta(\frac{Q_{\rm c}^2}{Q^2}-1)\quad (\rm if \quad Q<Q_{\rm c})\\
0\quad (\rm if \quad Q\geq Q_{\rm c}),
\label{GIalpha}
\end{array}
\right. 
\end{equation}
where $Q\equiv (c_{\rm s}\kappa_{\rm ep})/(\pi G \Sigma)$ is the Toomre's Q-value (\cite{ToomreQ}), $\kappa_{\rm ep}=\Omega$ is the epicyclic frequency for a Keplerian disk, $Q_{\rm c}$ is the critical Q-value below which gravitational torque acts, and $\eta~(<1)$ is the dimensionless number which represents the efficiency of angular momentum transport (Lin \& Pringle \yearcite{LP90}). 
According to the three-dimensional hydrodynamical calculations in Kratter et al. (\yearcite{Kratter}), 
$\alpha _{\rm G} \sim 1$ is indicated when the disk is gravitationally unstable. 
Boley et al. (\yearcite{Boley}) indicates that angular momentum transport by gravitational torque is important when $Q\leq 1.4$. 
In this paper, $Q_{\rm c}=1.4$ is adopted, and several cases with $\eta=10^{-3}-10^{-1}$ are considered. 

To calculate the temperature, local thermal equilibrium between the viscous heating rate $Q_{\rm visc}$ and radiative cooling rate $Q_{\rm cool}$ is assumed, i.e.,  
\begin{equation}
Q_{\rm visc}=Q_{\rm cool}. 
\label{radiatra}
\end{equation}
Each term is given by, 
\begin{equation}
Q_{\rm visc}=\frac{9}{8}\nu\Sigma (\frac{d\Omega}{dr})^2,\quad Q_{\rm cool}=\frac{32}{3}\frac{\sigma T^4}{\Sigma \kappa}. 
\label{QviscQcool}
\end{equation}
with Stephan-Boltzmann constant $\sigma$ and the opacity $\kappa$. 
For simplicity, we use the constant opacity $\kappa=3.0~{\rm cm^2 }~{\rm g}^{-1}$, which is the averaged value of $\kappa$ between several $10^2$ to $10^3~{\rm K}$ in Pollack et al. (\yearcite{PHB94}). 
Sound speed $c_{\rm s}$ and viscous parameter $\alpha$ are obtained when we solve the system of equations (\ref{alpha}) and (\ref{radiatra}), using the description of $\nu =\alpha c_{\rm s}h$, $Q=(c_{\rm s}\Omega)/(\pi G \Sigma)$, and $c_{\rm s}^2=(\gamma kT)/(\mu m_{\rm H})$, where $\gamma =7/5$ is the specific heat ratio, $k$ is the Boltzman's constant, $\mu =2.35$ is the molecular weight, and $m_{\rm H}$ is the mass of hydrogen. 
Equations (\ref{alpha}) and (\ref{radiatra}) constitute a non-linear set of equations with respect to $\alpha$ and $c_{\rm s}$. 
We solve $\alpha$ and $c_{\rm s}$ iteratively using Newton's method. 
In this section, equation (\ref{Mdotouter}) with $t_{\rm osci}=\infty$ is used as the outer boundary condition. 
Initial conditions, inner boundary conditions, and the grid number are the same as in section 2, except that boundary conditions for $c_{\rm s}$ is treated as $\partial c_{\rm s}/\partial r=0$ at $r=r_{\rm in}$ and $r=r_{\rm out}$. 
We investigate such features the interval and duration of outbursts driven by the GML, assuming that 
the mass accretion rate onto the inner disk is temporally constant. 
In Table 1, model parameters in this section are summarized. 
We terminate our calculation after outbursts occur a few dozen times or when the inner disk becomes a steady state without outburst. 
\begin{table}
  \caption{Parameters of equations in this paper}\label{tab:first}
  \begin{center}
    \begin{tabular}{c|c}
    \hline\hline
      $T_{\rm crit}$&$1,400~{\rm K}$\\
     $\alpha_{\rm M,on}$ &$10^{-2}$\\
     $\alpha_{\rm M,off}$ &$10^{-5}$\\
     $\kappa$ &$3.0~{{\rm cm}^2}~{\rm g}^{-1}$\\
      \hline
    \end{tabular}
  \end{center}
\end{table}
\subsection{Result}
We investigate the time variability of mass accretion rate at $r_{\rm in}$ for the several cases with different $\dot{M}_{\rm out}$. 
The time evolution of the mass accretion rate at $r_{\rm in}$ for several cases with constant $\dot{M}_{\rm out}(\simeq \dot{M}_0/2)$ and $\eta=10^{-2}$ are shown in Figure 2. 
\begin{figure}
\begin{center}
\FigureFile(80mm,80mm){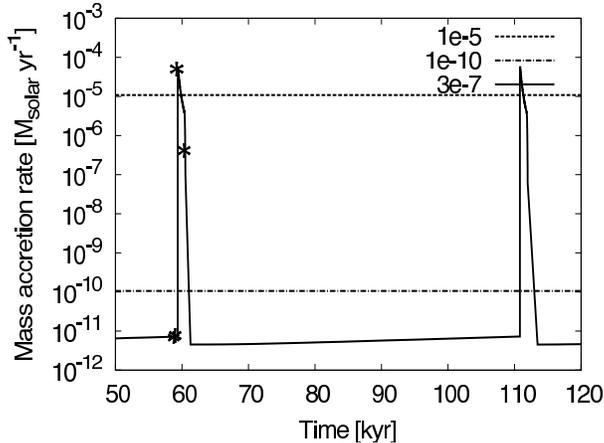}
\end{center}
\caption{Time variability of mass accretion rate at $r=r_{\rm in}$ for the case with $\dot{M}_{\rm out}=10^{-5}$, $10^{-10}$, and $3\times10^{-7}~{\rm M}_{\odot}~{\rm yr}^{-1}$.}
\end{figure} 
In Figure 2, it is seen that for the cases with $\dot{M}_{\rm out}=10^{-5}$ and $10^{-10} ~ {{\rm M}_{\odot} {\rm yr}^{-1}}$, the mass accretion rate at $r_{\rm in}$ is temporally constant without outburst in the late phase of calculation. 
	In Figure 2, only the period of $70~{\rm kyr}$ in the late phase is plotted for $\dot{M}_{\rm out}=10^{-10}$ and $10^{-5}$. 
At that phase, it is also found that $\dot{M}_{\rm in}=\dot{M}_{\rm out}$. 
In these two cases, the evolution is well described by a steady state model. 
In the case with $\dot{M}_{\rm out}=10^{-10} ~ {{\rm M}_{\odot} {\rm yr}^{-1}}$, $\alpha_{\rm M}$ stays equal to $\alpha_{\rm M,off}$ and $\alpha_{\rm G}$ equal to zero. 
This steady state with low $\dot{M}$ (LS state) appears when $\dot{M}_{\rm out}$ is sufficiently small. 
In the case with $\dot{M}_{\rm out}=10^{-5} ~ {{\rm M}_{\odot} {\rm yr}^{-1}}$, the disk is in a steady state with $\alpha_{\rm M}=\alpha_{\rm M,on}$ and $\alpha_{\rm G}=0$. 
This steady state with high $\dot{M}$ (HS state) appears when $\dot{M}_{\rm out}$ is sufficiently large. 
Using equations $(\ref{alpha})$, (\ref{radiatra}), and $\dot{M}\propto \nu \Sigma$, the relation between $Q$, $\alpha$, 
and $\dot{M}$ is given as $Q\propto {\dot{M}}^{-2/5}{\alpha}^{7/10}$.  
In the case with small $\dot{M}$, $Q$ is large and it is found that $Q>Q_{\rm c}$ for $\alpha=\alpha_{\rm M,off}=10^{-5}$ when $\dot{M}$ is smaller than $10^{-10}~ {{\rm M}_{\odot} {\rm yr}^{-1}}$. 
In the case with $\dot{M}_{\rm out}=10^{-5} ~ {{\rm M}_{\odot} {\rm yr}^{-1}}$, although $\dot{M}$ is $10^5$ times larger, $Q$ is only $10^{1/10}\sim 1.26$ times larger for $\alpha=\alpha_{\rm M,on}=10^{-2}=10^3\alpha_{\rm off}$. 
This makes $Q>Q_{\rm c}$ even in the case with $\dot{M}_{\rm out}=10^{-5}~ {{\rm M}_{\odot} {\rm yr}^{-1}}$. 
Thus, $\alpha_{\rm G}=0$ in both of LS and HS states. 

In Figure 2, it is also seen that outbursts occur in the case with $\dot{M}_{\rm out}=3\times 10^{-7}~ {{\rm M}_{\odot} {\rm yr}^{-1}}$. 
We set the time as $t=0$ when the first outburst starts to occur. 
In Figure 2, only $t>50~{\rm kyr}$ is shown because during $t<50~{\rm kyr}$ results are affected by the initial conditions. 
In Figure 2, it is seen that time variability of $\dot{M}_{\rm in}$ do not coincide with $\dot{M}_{\rm out}$. 
In this case, the inner disk is not in a steady state, 
and viscous parameters $\alpha_{\rm M}$ and $\alpha_{\rm G}$ are time dependent. 
In Figure 2, at $t\sim 60~{\rm kyr}$, the second outburst is seen. 
We pick up this outburst to study its features. 
Subsequent outbursts have the same features. 
Outbursts are characterized by duration of outburst $t_{\rm dur}=1.2\times 10^3~{\rm yr}$, the interval between outbursts $t_{\rm int}=5\times 10^4~\rm yr$, the maximum of mass accretion rate during outburst $\dot{M}_{\rm FU}=5.6\times 10^{-5}~ {{\rm M}_{\odot} {\rm yr}^{-1}}$, and mass accretion rate in quiescent phase between bursts $\dot{M}_{\rm TT}=(4-7)\times 10^{-12}~{{\rm M}_{\odot} {\rm yr}^{-1}}$. 
These characteristic values appear periodically until the end of our calculation at $10^6~{\rm yr}$. 
Note that $\dot{M}_{\rm FU}$ is larger than $\dot{M}_{\rm out}$ in these bursts. 
$\dot{M}_{\rm FU}$ is not driven by $\dot{M}_{\rm out}$ but driven by MRI with large $T>T_{\rm crit}$ and $\alpha_{\rm M,on}$. 
\begin{table}
  \caption{Conditions of spontaneous outburst to occur.}\label{tab:first}
  \begin{center}
    \begin{tabular}{c|c|c|c}
    \hline\hline
      $\dot{M}_{\rm out}~[{\rm M}_{\odot} {\rm yr}^{-1}]$&$\eta=10^{-3}$&$\eta=10^{-2}$&$\eta=10^{-1}$\\

      \hline
     $3\times10^{-6}\leqq\dot{M}_{\rm out}$ &\multicolumn{3}{c}{High $\dot{M}$ Steady state ({\rm HS state})}\\
     $10^{-10}<\dot{M}_{\rm out}<3\times 10^{-6}$ & \multicolumn{3}{c}{burst} \\
      $\dot{M}_{\rm out}\leqq10^{-10}$ &\multicolumn{3}{c}{Low $\dot{M}$ Steady state ({\rm LS state})}\\
      \hline
    \end{tabular}
  \end{center}
\end{table}

In addition, we investigate time variability of mass accretion rate $\dot{M}_{\rm in}$ at $r_{\rm in}$ for different values of $\eta=10^{-3}~{\rm and}~10^{-1}$. 
The results were similar. 
It is found that $t_{\rm dur}$, $\dot{M}_{\rm FU}$, and $\dot{M}_{\rm TT}$ are nearly independent of $\eta$, and the difference in $t_{\rm int}$ is within $3 \%$. 
It is also found that 
$\alpha_{\rm G}$ is a non-zero value and does not depend on $\eta$. 
When we use a small initial $\alpha_{\rm G}$ with $\eta$ smaller than $0.02$, angular momentum transport at the outer region of the disk is initially less efficient. 
Because $\dot{M}_{\rm out}$ is the same value, $\Sigma$ becomes larger and $Q$-value become smaller. 
The viscous parameter $\alpha_{\rm G}$ increases due to this disk evolution. 
On the other hand, when we use a larger $\eta$ and large initial $\alpha_{\rm G}$, angular momentum transport at the outer region of the disk becomes more efficient, and $\alpha_{\rm G}$ decreases. 
This behavior indicates that the value of $\alpha_{\rm G}$ with different $\eta$ is self-regulated by $Q$-value so that $\alpha_{\rm G}$ stays nearly the same (see also Takahashi et al. \yearcite{TIM13}). 
In Table 2, our results for several cases with $\dot{M}_{\rm out}$ and $\eta$ are summarized in a view point of whether outbursts of mass accretion rate occur or not in the inner disk. 
Whether outbursts occur or not are nearly independent of $\eta$, and strongly dependent of $\dot{M}_{\rm out}$. 
It is seen that outbursts occur in a limited range of mass accretion rates onto the inner disk $10^{-10}<\dot{M}_{\rm out}<3\times10^{-6}~{\rm M}_{\odot}~{\rm yr}^{-1}$. 
In addition, we calculate models with $\alpha_{\rm G}={\exp(-Q^4)}$ which is used in previous works (e.g., Takahashi et al. \yearcite{TIM13}, Zhu et al. \yearcite{Zhu09}, \yearcite{Zhu10a}, and \yearcite{Zhu10b} ) instead of equation (\ref{GIalpha}). 
We obtained almost the same range of $\dot{M}_{\rm out}$ for outburst. 
It is found that outbursts occur when $\alpha_{\rm G}$ has a value $\alpha_{\rm M,off}<\alpha_{\rm G}<\alpha_{\rm M,on}$. 
\begin{figure}
\begin{center}
\FigureFile(80mm,80mm){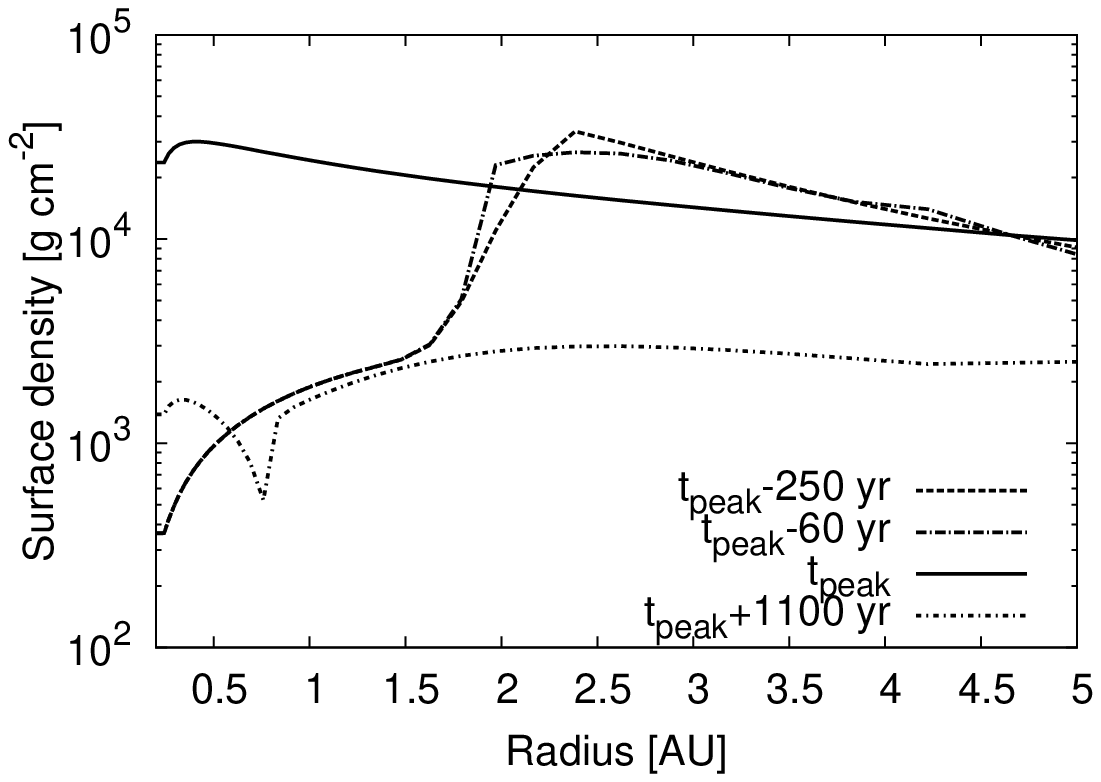}
\FigureFile(80mm,80mm){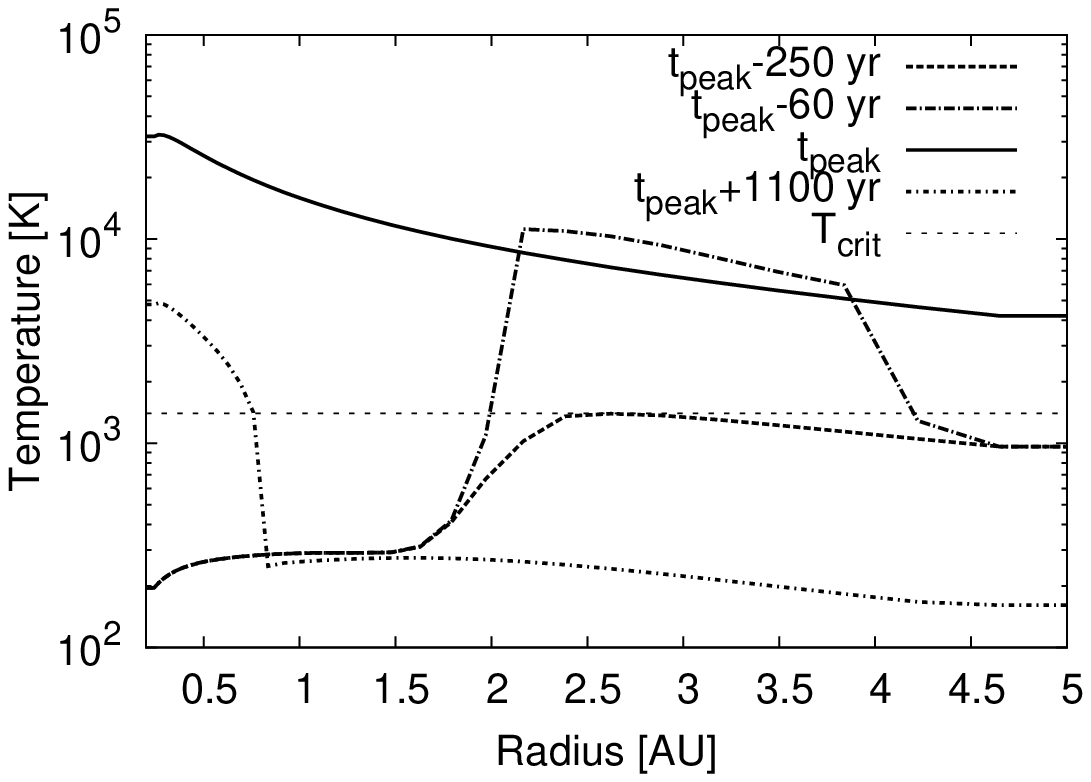}
\end{center}
\caption{Radial profiles of surface density $\Sigma$ (left diagram) and temperature $T$ (right diagram) in several epochs at $t_{\rm peak}-250~{\rm yr}$, $t_{\rm peak}-60~{\rm yr}$. $t_{\rm peak}$, and $t_{\rm peak}+1100~{\rm yr}$. }
\end{figure} 

In Figure 3, radial profiles of surface density $\Sigma$ and temperature $T$ in several epochs around $t_{\rm peak}=59,396~{\rm yr}$ during one burst are shown for the case with $\dot{M}_{\rm out}=3\times 10^{-7}~{\rm M}_{\odot} {\rm yr}^{-1}$. 
Each time instance in Figure 3 is indicated by the asterisk in Figure 2. 
In Figure 2, the asterisk at $t=t_{\rm peak}-250~{\rm yr}$ overlaps with that at $t=t_{\rm peak}-60~{\rm yr}$. 
At $t=t_{\rm peak}-250~{\rm yr}$, from Figure 2 it is seen that the inner disk is in a quiescent state. 
In the right diagram of Figure 3, it is seen that $T<T_{\rm crit}$ and the inner disk is in the MRI-inactive state. 
However, the matter is pilling up due to mismatch in the mass accretion rate between the inner region ($r<2.5~{\rm AU}$) and the outer region ($r>2.5~{\rm AU}$), 
and temperature at $r=2.5~{\rm AU}$ is increasing. 
At $t=t_{\rm peak}-60~{\rm yr}$, 
it is seen that the region with $T>T_{\rm crit}$ and $\alpha =\alpha_{\rm M,on}=10^{-2}$ (MRI active region) appears. 
After the MRI active region with $\alpha_{\rm M, on}$ appears at $r=2.5~{\rm AU}$,  
rapid accretion to the inner region and diffusion to the outer region occur, 
since angular momentum transport in the active region is more efficient than the other region. 
Surface density and temperature of the outer ($r\gtrsim4.5~{\rm AU}$) and the inner region ($r\lesssim 2~{\rm AU}$) increase due to this mass redistribution. 
The MRI active region ($T>T_{\rm crit}$) propagates both to the inner and the outer region of the inner disk.  
At $t=t_{\rm peak}$, the MRI active region reach the center and the wholes disk becomes MRI active. 
Indeed, both of $\Sigma$ and $T$ is high due to mass redistribution from MRI active region around $r=2.5~{\rm AU}$. 
In Figure 2, it is seen that $\dot{M}_{\rm in}=5.6\times 10^{-7} ~{\rm M}_{\odot} {\rm yr}^{-1}$ at $t=t_{\rm peak}$ is higher than $\dot{M}_{\rm out}$. 
The matter in the inner disk is accreted onto the star faster than it is replenished due to accretion from outer region of the inner disk, 
and surface density and temperature in the inner disk start to decrease. 
As a result, at $t_{\rm peak}+1100~{\rm yr}$, a large amount of mass is accreted, and $\Sigma$ becomes low. 
Temperature is lower than $T_{\rm crit}$, and thus the inner disk is in the MRI inactive state. 

In Figure 4, the time evolution of mass accretion rate at $r_{\rm in}$ for the case with $\dot{M}_{\rm out}=1\times 10^{-6}$ and $3\times 10^{-7}~{\rm M}_{\odot} {\rm yr}^{-1}$ is shown. 
\begin{figure}
\begin{center}
\FigureFile(80mm,80mm){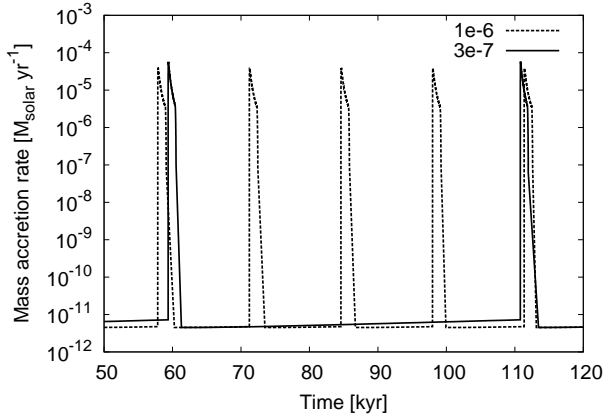}
\end{center}
\caption{Time variability of mass accretion rate at $r_{\rm in}$ for several cases with $\dot{M}_{\rm out}$. }
\end{figure} 
It is seen that the maximum of $\dot{M}$ at $r_{\rm in}$ ($\dot{M}_{\rm FU}$) and duration of outburst $t_{\rm dur}$ are almost the same in two cases. 
Oppositly, the interval between the outbursts is quite different. 
\begin{table}
  \caption{Difference in the interval between outbursts}\label{tab:first}
  \begin{center}
    \begin{tabular}{|c|c|}
    \hline\hline
     $\dot{M}_{\rm out} ~[{\rm M}_{\odot} {\rm yr}^{-1}]$& interval [yr]\\
     \hline
     $1\times 10^{-6}$&$1.2\times10^4$\\
      \hline
    $3\times 10^{-7}$&$5\times 10^4$\\
          \hline
    \end{tabular}
  \end{center}
\end{table}
Differences in the interval for two cases are summarized in Table 3. 
It is seen that the interval $t_{\rm int}$ for the case with $\dot{M}_{\rm out}=1\times 10^{-6}~{\rm M}_{\odot} {\rm yr}^{-1}$ is shorter than that for $\dot{M}_{\rm out}=3\times 10^{-7}~{\rm M}_{\odot} {\rm yr}^{-1}$. 
This is because it takes a longer time to replenish the disk with matter until $T>T_{\rm crit}$ in the case with $\dot{M}_{\rm out}=3\times 10^{-7}~{\rm M}_{\odot} {\rm yr}^{-1}$, than in the case with $\dot{M}_{\rm out}=1\times 10^{-6}~{\rm M}_{\odot} {\rm yr}^{-1}$. 

In summary, in this section it is shown that the spontaneous outburst occurs when $10^{-10}<\dot{M}_{\rm out}<3\times10^{-6}~{\rm M}_{\odot} {\rm yr}^{-1}$ and that the value of $\dot{M}_{\rm out}$ affects the interval between the bursts $t_{\rm int}$ but not the value of $\dot{M}_{\rm in}$. 
Note that our results suggest that time variability of $\dot{M}_{\rm in}$ is possible without time variability of $\dot{M}_{\rm out}$, and that features of outburst, i.e., $t_{\rm dur}$, $t_{\rm int}$, $\dot{M}_{\rm FU}$, and $\dot{M}_{\rm TT}$ are nearly independent of $\dot{M}_{\rm out}$ except for $t_{\rm int}$. 
$\dot{M}_{\rm FU}$ is larger than $\dot{M}_{\rm out}$. 
The low boundary for the steady accretion rate sensitively depends on the $\alpha_{\rm M,off}$, which we know little about. 

\section{Combination of Episodic Accretion and GML} 
In this section, we investigate the time variability of accretion rate at $r_{\rm in}$ due to a combination of episodic secretion imposed onto disk's outer edge and the GML operating within the disk. 
Viscous evolution between $r_{\rm in}$ and $r_{\rm out}$ is calculated by taking into account $\alpha_{\rm G}$ and $\alpha_{\rm M}$ as well as episodic accretion rate onto the inner disk $\dot{M}_{\rm out}(t)$ by solving equations (\ref{diffusiveeq}), (\ref{alpha}), and (\ref{radiatra}) with parameters given in Table 1. 
For the pattern of episodic accretion, we consider two types of $\dot{M}_{\rm out}(t)$. 
One is based on an analytic formula, and the other is based on numerical hydrodynamical calculations. 
Inner boundary condition is the same as in section 2 and 3.  

\subsection{Analytic $\dot{M}_{\rm out}(t)$ with a Single Period}
In this subsection, as the outer boundary condition, we use the same formula as equation (\ref{Mdotouter}) in section 2.1. 
In equation (\ref{Mdotouter}), we use $\eta=10^{-2}$, $\dot{M}_{0}=2\times 10^{-8}~{\rm M}_{\odot} {\rm yr}^{-1}$, and $\delta =2\times 10^{-3}$. 
By this combination of $\dot{M}_{0}$ and $\delta$, the mean accretion rate in a single oscillation period is given by $<\dot{M}_{\rm out}(t)>_{\rm osci}=3\times 10^{-7}~{\rm M}_{\odot}{\rm yr}^{-1}$, which is the same as the value of constant $\dot{M}_{\rm out}(t)$ in section 3. 
We use a finite $t_{\rm osci}$, the value of which is different from that of section 3. 

\begin{figure}
\begin{center}
\FigureFile(80mm,80mm){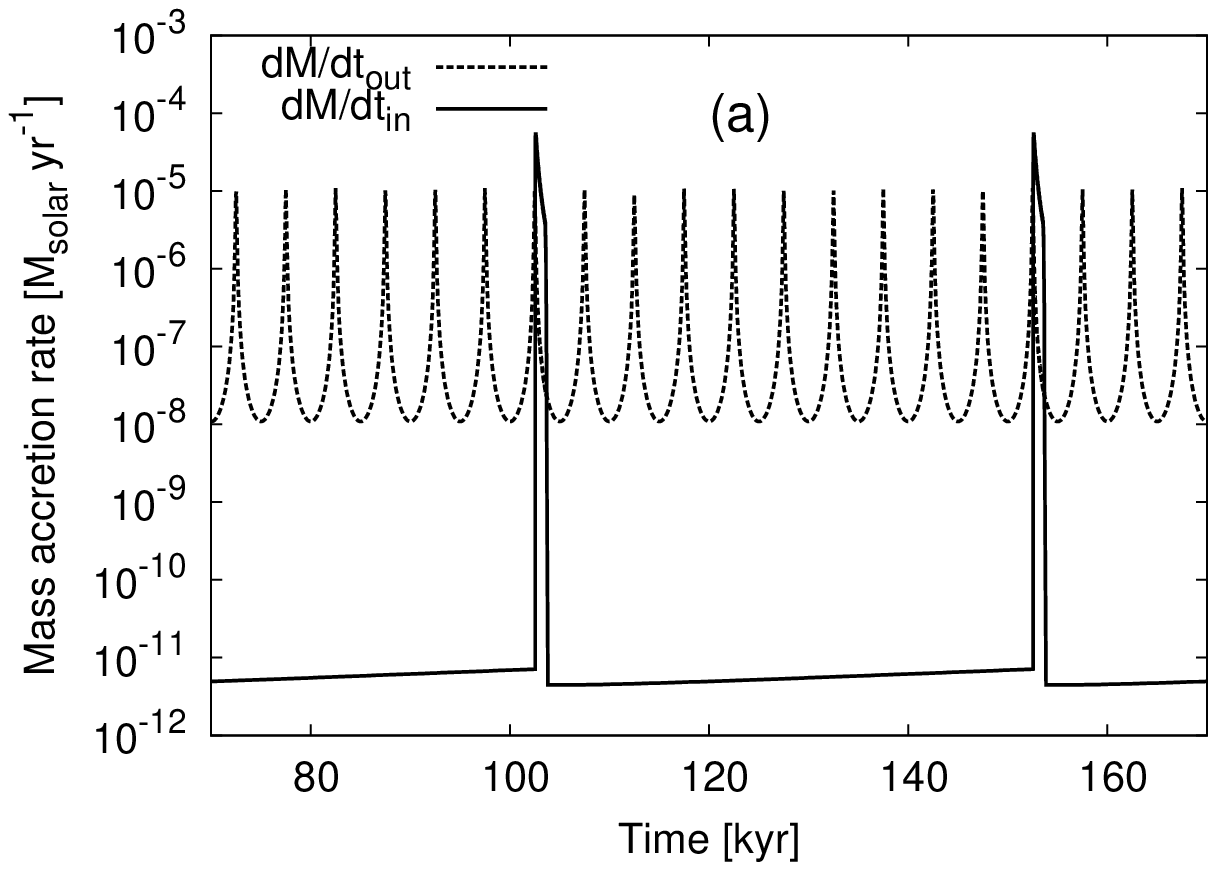}
\FigureFile(80mm,80mm){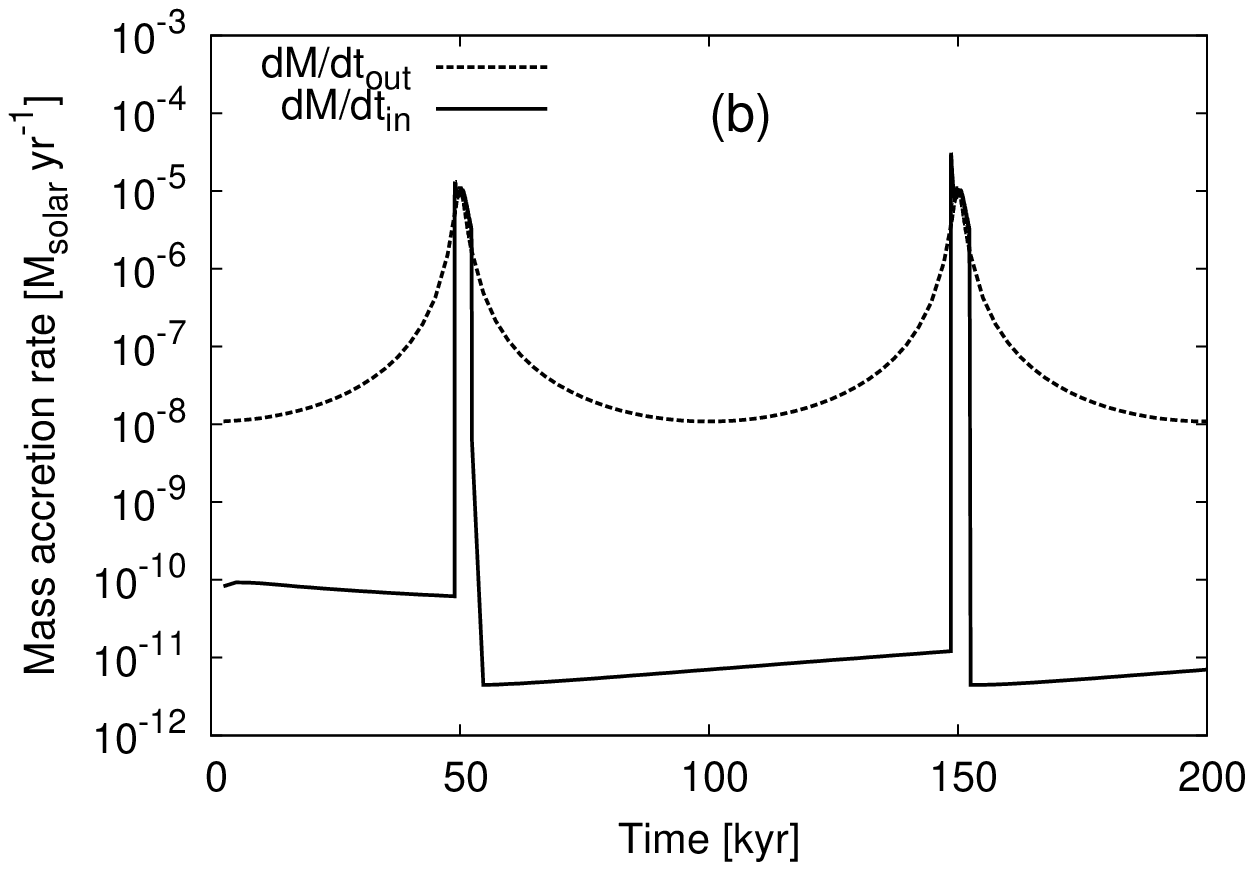}
\end{center}
\caption{Time variability of mass accretion rate at $r_{\rm out}$ and $r_{\rm in}$ for the case with $t_{\rm osci}=5~{\rm kyr}$ (a) and $t_{\rm osci}=100~{\rm kyr}$ (b). }
\end{figure} 

In Figure 5 (a) and (b), the time evolution of the mass accretion rate at $r_{\rm out}$ and $r_{\rm in}$ for the cases with $t_{\rm osci} =5~{\rm kyr}$ and $t_{\rm osci} =100~{\rm kyr}$ are shown, respectively. 
In Figure 5 (a), it is seen that $\dot{M}_{\rm in}$ (solid line) do not coincide with $\dot{M}_{\rm out}$ (dotted line) and that outbursts of $\dot{M}_{\rm in}$ occur at $t=105~{\rm kyr}$ and $t=155~{\rm kyr}$. 
This is because the inner disk acts to damp the oscillation in $\dot{M}_{\rm out}$ when transporting the accreted matter from the outer to the inner boundary, as explained in section 2.2 and because the GML drives outbursts. 
The outbursts have the following properties: $t_{\rm dur}=1.2\times 10^3 \rm yr$, $t_{\rm int}=5\times 10^4~\rm yr$, $\dot{M}_{\rm FU}=5.6\times 10^{-5}~ {{\rm M}_{\odot} {\rm yr}^{-1}}>\dot{M}_{\rm out}$, and $\dot{M}_{\rm TT}=(4-7)\times 10^{-12}~{{\rm M}_{\odot} {\rm yr}^{-1}}$.  
These values are similar to the case with constant $\dot{M}_{\rm out}=3\times 10^{-7}~{\rm M}_{\odot}{\rm yr}^{-1}$ in Figure 4 in section 3. 
By comparing Figure 4 and Figure 5 (a), it is found that the resulting $\dot{M}_{\rm in}$ are similar although $\dot{M}_{\rm out}$ in Figure 4 is temporally constant and $\dot{M}_{\rm out}$ in Figure 5 (a) includes time dependence. 
Thus, outbursts in Figure 5 (a) can be regarded as the same mode as in section 3. 
Since this mode can be reconstructed by the GML with constant $\dot{M}_{\rm out}$, 
we call this a spontaneous mode of outburst. 
It is also found that this spontaneous mode of outburst appears only when $t_{\rm osci}<t_{\rm {int, c}}$, where $t_{\rm {int, c}}$ is the interval between outbursts in the case with constant $\dot{M}_{\rm out}=<\dot{M}_{\rm out}>_{\rm osci}$. 
Here, note that $t_{\rm osci}$ is compared with $t_{\rm {int,c}}$ instead of $t_{\rm diff}$ in section 2.2 because $t_{\rm diff}$ is unuseful time-dependent quantity here. 

For the case with $t_{\rm osci} =100 {\rm kyr}>t_{\rm int,c}$ in Figure 5 (b), it is seen that the period of $\dot{M}_{\rm in}$ (solid line) is the same as the period of $\dot{M}_{\rm out}$ (dotted line), although $\dot{M}_{\rm in}$ is different from $\dot{M}_{\rm out}$. 
It is also seen that $\dot{M}_{\rm in}$ has features of outburst $t_{\rm dur}\sim 3.5\times10^3~{\rm yr}$, $t_{\rm int}=10^5~\rm yr$, $\dot{M}_{\rm FU}=1\times 10^{-5}~ {{\rm M}_{\odot} {\rm yr}^{-1}}$, and $\dot{M}_{\rm TT}=(4-10)\times 10^{-12}~{{\rm M}_{\odot} {\rm yr}^{-1}}$. 
These are different from features in Figure 4 and 5 (a), i.e., $t_{\rm dur}$ in Figure 5 (b) is 3 times longer, $t_{\rm int}$ is 2 times longer, $\dot{M}_{\rm FU}$ is 5 times smaller, and range of $\dot{M}_{\rm TT}$ is twice broader than in Figure 4 and 5 (a), respectively. 
Thus, outbursts in Figure 5 (b) should be regarded a mode different from those of section 3 and Figure 5 (a). 
In this new mode, time-dependent $\dot{M}_{\rm out}$ affects $\dot{M}_{\rm in}(t)$. 
We call this a stimulated mode of outburst. 
Futhermore, in Figure 5 (b), it is seen that at the burst phase at $t\sim 50 {\rm kyr}$, $\dot{M}_{\rm in}$ roughly coincides with $\dot{M}_{\rm out}$, i.e., $\dot{M}_{\rm in}\sim \dot{M}_{\rm out}\sim 10^{-5}~(>3\times 10^{-6})~{\rm M}_{\odot} {\rm yr}^{-1}$. 
This rough agreement indicates that the inner disk can be approximated to be in a quasi-steady state (HS state in Table 2) caused by active MRI. 
The agreement between $\dot{M}_{\rm in}$ and $\dot{M}_{\rm out}$ at and near the burst phase occur when roughly constant $\dot{M}_{\rm out}>3\times 10^{-6}~{\rm M}_{\odot} {\rm yr}^{-1}$ ($\dot{M}_{\rm out}$ for HS state) is kept longer time than the diffusion time $t_{\rm diff}$ in the MRI active state (several $10^3~{\rm yr}$) because information at $r_{\rm out}$ reach $r_{\rm in}$ in $t_{\rm diff}$. 

In summary, from Figure 5, it is shown that the GML can drive outbursts even if the oscillation of $\dot{M}$ decays during transmitting the inner disk. 
We identified two different modes of outburst which are spontaneous one and stimulated one. 

\subsection{Numerical $\dot{M}_{\rm out}$ produced by Hydrodynamic Simulations}
In the last subsection 4.1, we used analytic formula (equation (\ref{Mdotouter})) for $\dot{M}_{\rm out}(t)$. 
In this subsection, we use a more realistic $\dot{M}_{\rm out}(t)$ as the outer boundary conditions in our model. 
Equations (\ref{diffusiveeq}), (\ref{alpha}), and (\ref{radiatra}) are calculated with $\dot{M}_{\rm out}$ which is taken from two-dimensional numerical hydrodynamics simulation of 
Vorobyov \& Basu (\yearcite{VB10}).  
They calculated the formation and global evolution of the circumstellar disk and dynamically in falling envelope as well as the central protostar (a modification presented in Vorobyov et al. \yearcite{vor13}), using the thin-disk approximation. 
We used the mass accretion rate through the sink cell at $5\rm AU$ from the central star during $0.3<t<0.5~{\rm Myr}$ after beginning of the cloud core collapse. 
The mass of the central star in hydrodynamical simulations is $M_{*}=0.44-0.53~{\rm M}_{\odot}$ during this time. 
We approximate it by a constant value $M_*=0.5~{\rm M}_{\odot}$ in our calculation. 

In Figure 6, the time evolution of the mass accretion rate at $r_{\rm out}$ and $r_{\rm in}$ is shown. 
\begin{figure}
\begin{center}
\FigureFile(80mm,80mm){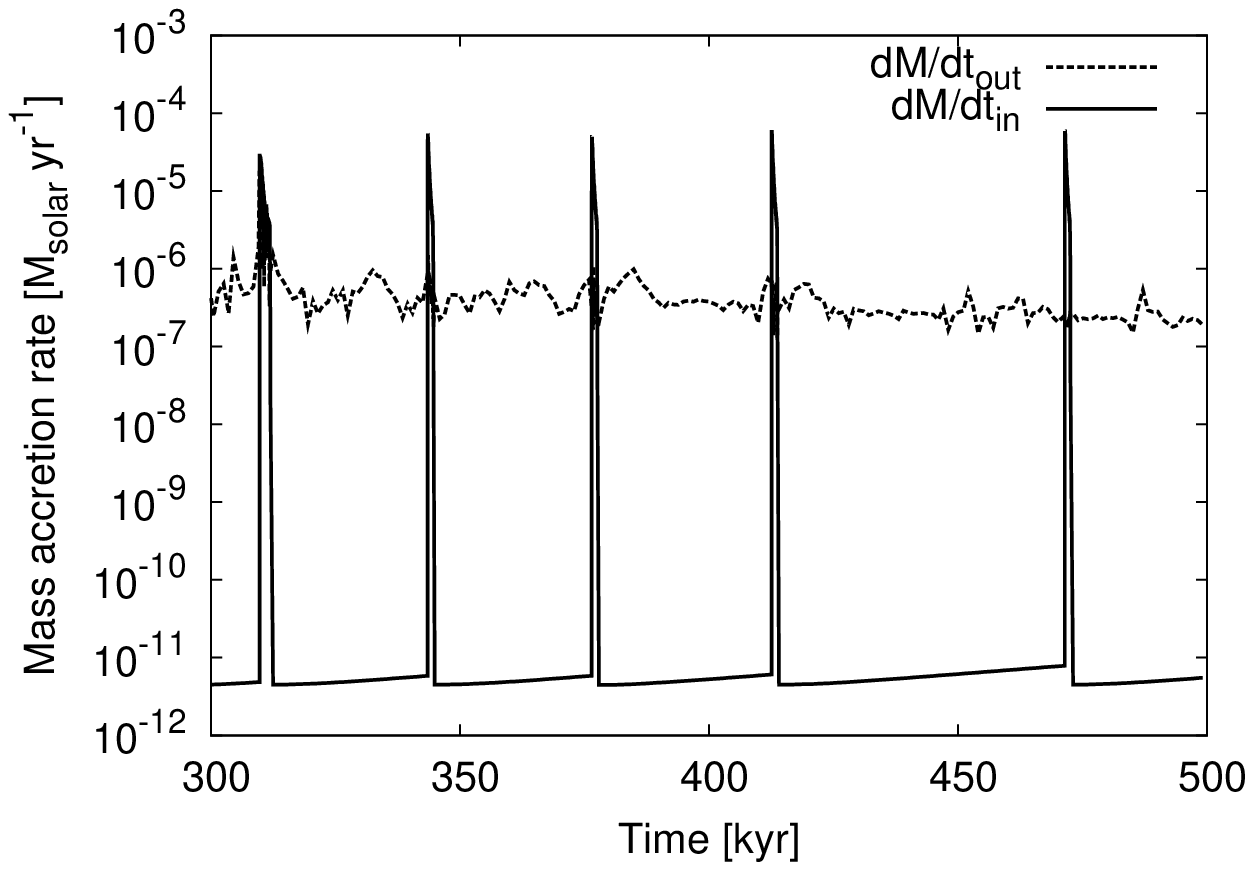}
\FigureFile(80mm,80mm){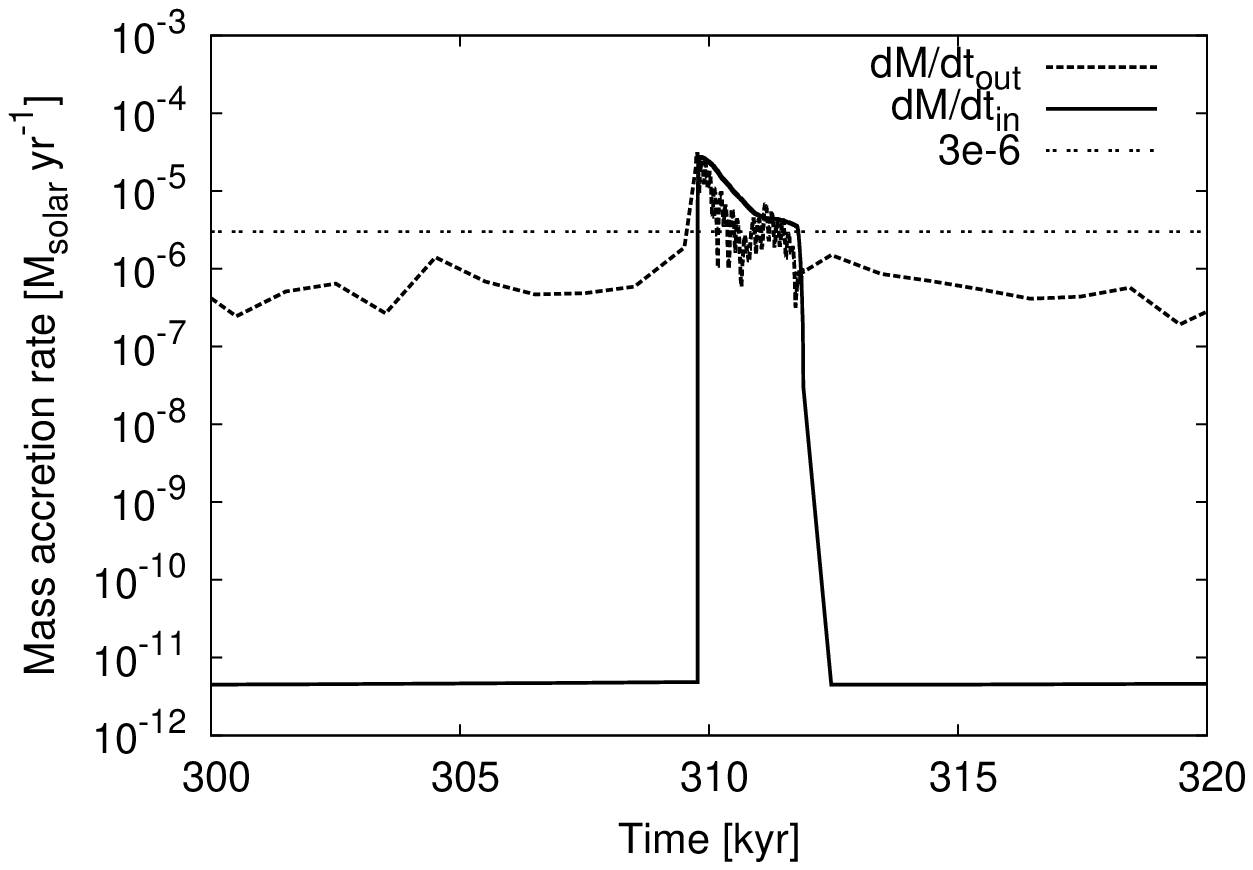}
\end{center}
\caption{Time variability of mass accretion rate at $r_{\rm out}$ and $r_{\rm in}$ for the case with $\dot{M}_{\rm out}$ by hydrodynamical simulation (see sentences for more detail.) }
\end{figure} 
It is seen that the time behavior of $\dot{M}_{\rm out}$ (dotted line) is complicated but the typical time scale of fluctuation is shorter than several kyr. 
The time averaged value of $\dot{M}_{\rm out}(t)$ during several ${\rm kyr}$ is gradually decreasing. 
For example, $<\dot{M}_{\rm out}>\sim 5\times 10^{-7}~{\rm M}_{\odot}{\rm yr}^{-1}$, and $3\times 10^{-7}~{\rm M}_{\odot}{\rm yr}^{-1}$ during $315<t<415~{\rm kyr}$, and $415<t<471~{\rm kyr}$, respectively. 
During $315<t<500~{\rm kyr}$, $\dot{M}_{\rm out}$ fluctuates around several $10^{-7}~{\rm M}_{\odot}{\rm yr}^{-1}$ and does not exceed $3\times 10^{-6}~{\rm M}_{\odot}{\rm yr}^{-1}$. 
Conversely, in the right diagram of Figure 6, it is seen that $\dot{M}_{\rm out}$ has a burst with a maximum value of a few $\times 10^{-5}~{\rm M}_{\odot}{\rm yr}^{-1}$ at around $310~{\rm kyr}$ 
and $<\dot{M}_{\rm out}>$ is about $3\times 10^{-6}~{\rm M}_{\odot}{\rm yr}^{-1}$ during $310<t<312~{\rm kyr}$. 
In Figure 6, many outbursts are seen in $\dot{M}_{\rm in}$ (solid line) even with the realistic $\dot{M}_{\rm out}(t)$. 
It is seen that the time behavior of $\dot{M}_{\rm in}$ generally differs from that of $\dot{M}_{\rm out}$. 
However, $\dot{M}_{\rm in}$ has values of about $10^{-6}-10^{-5} ~{\rm M}_{\odot} {\rm yr}^{-1}$, similar to those of $\dot{M}_{\rm out}$ when the state with $\dot{M}_{\rm out}>10^{-6}~{\rm M}_{\odot} {\rm yr}^{-1}$ is kept at $t=310-312~{\rm kyr}$. 
$\dot{M}_{\rm in}$ in this burst remarkably coincides with $\dot{M}_{\rm out}$ (see Figure 5 (b)) which is the stimulated mode of outburst discussed in section 4.1. 
In the phase during $415<t<471 ~{\rm kyr}$, it is seen that the interval between outbursts is about $5\times 10^4 {\rm yr}$, which is almost the same as that in Figure 5 (a). 
In the phase during $315<t<415 ~{\rm kyr}$, it is seen that the interval between outbursts becomes shorter, about $3\times 10^4 {\rm yr}$. 
This is because the time-averaged value of $\dot{M}_{\rm out}$ during $315<t<415 ~{\rm kyr}$ is larger than that during $415<t<471 ~{\rm kyr}$. 
The peak of $\dot{M}_{\rm in}$ is larger than $\dot{M}_{\rm out}$ in spontaneous outbursts. 
Outbursts at $t\sim 345,~375,~415,~{\rm and}~465~{\rm kyr}$ belong to the spontaneous mode of outburst discussed in section 4.1 because $\dot{M}_{\rm out}$ is smaller than $3\times 10^{-6}~{\rm M}_{\odot} {\rm yr}^{-1}$. 

In summary of section 4.2, we investigated the time variability of $\dot{M}_{\rm in}$ using realistic $\dot{M}_{\rm out}$. 
Our results in Figure 6 confirm our findings in section 2, 3, and 4.1 in the sense that outbursts are driven by the GML although the fluctuation in $\dot{M}$ may decay when passing through the inner disk and that two modes of outburst (spontaneous one and stimulated one) are proved to exist in the case with realistic $\dot{M}_{\rm out}$.  

\subsection{Effect of Full Opacity Table}
Finally, we consider the effect of full opacity table $\kappa$. 
In this subsection, we use the Rosseland mean opacity, $\kappa(\rho, T)$, which is approximated by using power-law as 

\begin{equation}
\kappa=\kappa_0\rho^{a}T^{b},
\label{BellLinkappa}
\end{equation}
where $\rho$ is volume density and the values of $\kappa_0$, $a$, and $b$ are summarized in Table 4 (see also Bell \& Lin \yearcite{BellLin1994}, Cossins et al. \yearcite{Cos10}, and Kimura \& Tsuribe \yearcite{kim12}).

\begin{table}
  \caption{Bell and Lin opacity}\label{tab:comp_obs}
  \begin{center}
    \begin{tabular}{|cccccc|}
    \hline
      Opacity regime &$\kappa_0 ({\rm cm}^2{\rm g}^{-1})$&a&b&Temperature from (K)&Temperature to (K) \\
     \hline
      Ices&$2\times 10^{-4}$&$0$&$2$&$0$&$166.8$\\
     \hline
      Sublimation of ices&$2\times10^{16}$&$0$&$-7$&$166.8$&$202.6$ \\
      \hline
      Metal dust&$0.1$&$0$&$1/2$&$202.6$&$2286.7\rho^{2/49}$\\
          \hline
      Sublimation of metal dust&$2\times 10^{81}$&$1$&$-24$&$2286.7\rho^{2/49}$&$2029.7\rho^{1/81}$\\
          \hline
    Molecules&$3\times 10^{-5}$&$2/3$&$3$&$2029.7\rho^{1/81}$&$10000\rho^{1/21}$\\
	\hline
     Hydrogen scattering&$1\times 10^{-36}$&$1/3$&$10$&$10000\rho^{1/21}$&$31195.2\rho^{4/75}$\\
     \hline
     Bound-free and free free&$1.5\times 10^{-20}$&$1$&$-5/2$&$31195.2\rho^{4/75}$&\\
      \hline
          
    \end{tabular}
  \end{center}
\end{table}

We use the energy equation instead of equation (\ref{QviscQcool}) as,
\begin{equation}
\frac{dE}{dt}=\frac{1}{\Sigma}(Q_{\rm visc}-Q_{\rm cool}),
\label{equationconserv}
\end{equation}
where $E=c_{\rm s}^2/\gamma(\gamma-1)$ is the specfic internal energy. 
Different from the previous sections, we use radiative cooling rate as 
\begin{equation}
Q_{\rm cool}=\frac{32}{3}\frac{\sigma T^4}{\Sigma \kappa} (\kappa \Sigma>1), \quad
\frac{8\sigma T^4 \Sigma \kappa}{3}(\kappa \Sigma<1), 
\label{coolingratelowtau}
\end{equation}
because the inner disk may become optical thin.   
In order to see effects of different expression of $\alpha_{\rm G}$, we use 
\begin{equation}
\alpha_{\rm G}={\exp(-Q^4)}.
\label{ZHalphaG}
\end{equation}

Equations (\ref{diffusiveeq}), (\ref{alpha}), (\ref{BellLinkappa}), (\ref{equationconserv}), and (\ref{ZHalphaG}) are calculated with realistic $\dot{M}_{\rm out}$ which is the same value used in section 4.2. 
As the test case, we calculated these equations with constant $\dot{M}_{\rm out}$ and $T_{\rm crit}=800{\rm K}$, which is the critical temperature used in Armitage et al. (\yearcite{Armitage01}). 
We found that outbursts occur in a range of $10^{-9}<\dot{M}_{\rm out}< 3\times 10^{-6}~{\rm M}_{\odot} {\rm yr}^{-1}$ for the case with $T_{\rm crit}=800{\rm K}$. 
We confirmed that the critical mass accretion rate for achieving a steady accretion is almost the same as in the previous study (Armitage et al. \yearcite{Armitage01}). 

\begin{figure}
\begin{center}
\FigureFile(80mm,80mm){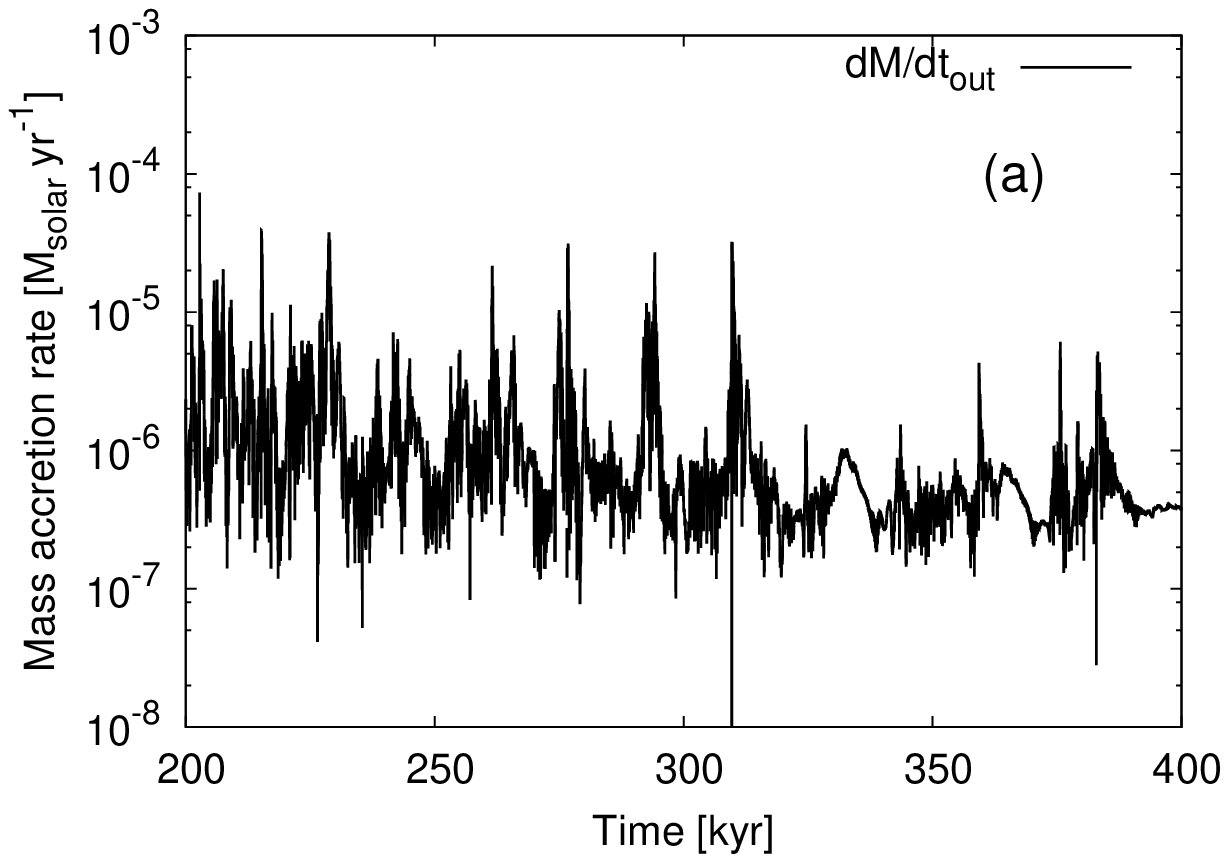}
\FigureFile(80mm,80mm){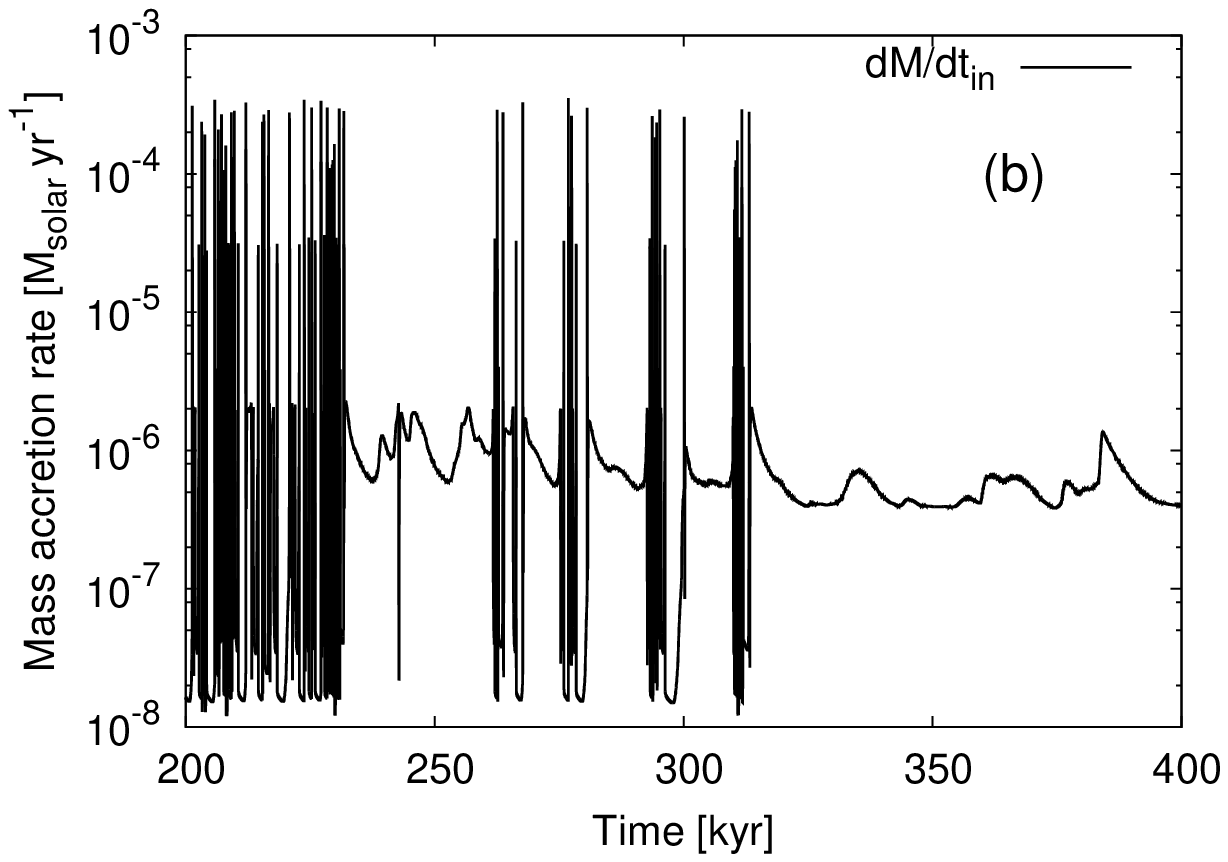}\\
\FigureFile(80mm,80mm){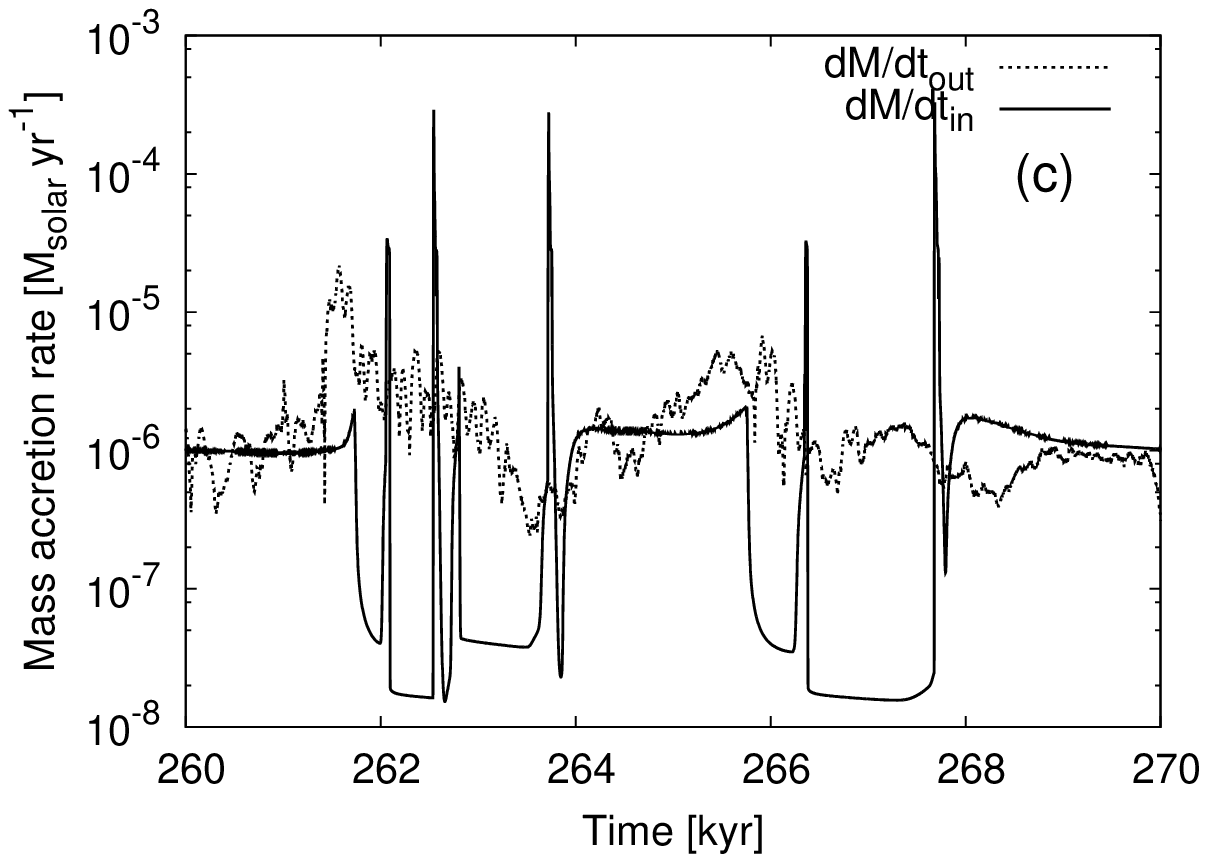}
\end{center}
\caption{Time variability of mass accretion rate at $r_{\rm out}$ and $r_{\rm in}$ for the case with full opacity table (see sentences for more detail.)}
\end{figure} 
In Figure 7, time evolutions of mass accretion rate at $r_{\rm out}$ (a) and $r_{\rm in}$ (b) are shown. 
It is seen that outbursts occur even in the case with full opacity table $\kappa(\rho, T)$. 
Different from section 3.2, 4.1, and 4.2, it is seen that bursts have shorter time scale. 
$\dot{M}_{\rm TT}$ is among $10^{-8}-10^{-7}~{\rm M}_{\odot}~{\rm yr}^{-1}$, which is about $10^3$ times larger, and $t_{\rm dur}\sim10^2~{\rm yr}$, which is $10$ times shorter than the result in the previous sections, respectively. 
This is because opacity $\kappa$ becomes much smaller than constant $\kappa=3$ when $T>T_{\rm crit}$. 
Radiative cooling rate $Q_{\rm cool}$ during the burst becomes larger than the case with constant $\kappa =3$. 
The burst continues shorter, and smaller amount of mass is accreted during the burst than the case with constant $\kappa=3$. 
From equations (\ref{radiatra}) and (\ref{QviscQcool}), $\Sigma\propto \kappa^{-1/2}$ when $T=T_{\rm crit}$, 
thus surface density $\Sigma$ after the burst is larger than the case with constant $\kappa=3$. 
We checked that the viscous parameter $\nu=\alpha {c_{\rm s}}^2/\Omega$ is almost the same in each case because temperature after the burst is almost the same. 
Thus, $\dot{M}_{\rm TT}\sim\nu\Sigma$ becomes larger. 

It is found that the behavior of $\dot{M}_{\rm in}$ becomes more complicated than the case with constant $\kappa=3$ in the previous sections. 
However, we found that the results are qualitatively consistent. 
From Figure 7 a, b, and c, it is seen that the time behavior of $\dot{M}_{\rm in}$ does not coincide with $\dot{M}_{\rm out}$. 
From figure 7 c, it is also seen that bursts of $\dot{M}_{\rm in}$ occur about $1 \rm kyr$ after $\dot{M}_{\rm out}$ reaches $10^{-5}~{\rm M}_{\odot}~{\rm yr}^{-1}$.   
The mass accretion rate onto the disk $\dot{M}_{\rm out}$ does not appear $\dot{M}_{\rm in}$ directly. 
Large amount of $\dot{M}_{\rm out}$ triggers the spontaneous outbursts in the inner disk.

\section{Discussions}
\subsection{Comparison with the previous theoretical studies}

Some hydrodynamical simulations about formation of a disk by using sink cells 
claim that FU Ori-type outbursts are explained by disk fragmentation \citep{VB06}. 
The mass accretion rate onto sink radii $\dot M_{\rm sink}$ is strongly fluctuated in their simulation. 
Although they assume $\dot M_{\rm sink}$ to be equivalent to the mass accretion rate onto protostar $\dot M_*$, 
this assumption needs justification because they set sink cells whose radii are too large. 
We show that fluctuations of $\dot M_{\rm sink}$ may decay during passing through the inner disk 
when the diffusion time of inner disk is much larger than the time scale for oscillation of fluctuations (see section 2). 
This result clearly indicates that it is important to consider an inner disk within the sink cell. 
Our modeling also shows that a sharp increase in disk accretion at around 5 AU, 
caused for example by a massive clump approaching to and disintegrating near the star, 
can trigger a true mass accretion burst onto the star by pushing the inner disk into the MRI-active state.

Some researches show that outbursts like FU orionis can be explained by the GML model 
using one-dimensional disk models with realistic opacities 
and realistic mechanisms of angular momentum transport \citep{Armitage01,Zhu09,ML11}. 
However, their models cannot include the fluctuation in the outer mass accretion rate 
induced by fragmentation of the disk. 
We calculate the mass accretion rate onto a protostar with the fluctuated outer mass accretion rate, 
and newly find that fluctuations of $\dot M_{\rm out}$ affect the $t_{\rm int}$ rather than the value of $\dot M_{\rm in}$ itself. 
It is also found that $t_{\rm int}$ is smaller for larger $\dot M_{\rm out}$ owing to rapid mass loading from the outer region, 
These results show that fluctuations of $\dot M_{\rm out}$ has non-negligible influence on mass accretion rate onto the protostar. 
Multi-dimensional calculation is necessary in order to discuss property of outbursts further in detail. 

\subsection{Comparison with observations}
Constraints from observations about FU Ori-type outbursts are shown in Table \ref{tab:comp_obs}
(see also \cite{BellLin1994}, \cite{sta04}, \cite{aud14}). 
Mass accretion rates in FU Ori stage are about
$\dot M_{\rm FU}\sim 10^{-5}-10^{-4}~{\rm M}_{\odot}~{\rm yr}^{-1}$, 
which are roughly $10^2-10^3$ times higher than those in T-Tauri phase, 
$\dot M_{\rm TT}\sim 10^{-8}-10^{-7}~{\rm M}_{\odot}~{\rm yr}^{-1}$. 
Time scales of the duration of outburst $t_{\rm dur}$ are estimated by decreasing rate of luminosity as 
$t_{\rm dur}\sim$ a few $\times$ 10-100 year. 
Time scale of interval between FU Orionis events $t_{\rm int}$ is difficult to estimate 
because it cannot be directly observed. 
It is estimated statistically as $10^3-10^5~{\rm yr}$. 

The results in our model is also tabulated in Table \ref{tab:comp_obs}. 
These values are obtained with the parameter tabulated in Table 1. 
Mass accretion rate onto the star in the T-Tauri phase and in the outburst phase 
are around $\dot M_{\rm TT}\sim 10^{-12}~{\rm M}_{\odot}~{\rm yr}^{-1}$ 
and $\dot M_{\rm FU} \sim 10^{-5}~{\rm M}_{\odot}~{\rm yr}^{-1}$, respectively. 
Note that these mass accretion rate are independent of the outer mass accretion rate $\dot M_{\rm out}$
(see section 3 for more detail). 
The duration time of an outburst in our result is about $t_{\rm dur}\simeq 1\times 10^3 ~{\rm yr}$, 
which is also nearly independent of $\dot M_{\rm out}$. 
The time interval between outbursts is about $10^4~{\rm yr}$ and $10^6~{\rm yr}$ 
for the case with $\dot M_{\rm out}=10^{-6}~{\rm M}_{\odot}~{\rm yr}^{-1}$ 
and $\dot M_{\rm out}=10^{-8}~{\rm M}_{\odot}~{\rm yr}^{-1}$, respectively.  
Although $t_{\rm int}$, and $\dot M_{\rm FU}$ in our model 
are not so different from those of observations, 
$\dot M_{\rm TT}$ and $t_{\rm int}$ in our model 
is much smaller and longer than those of observations, respectively. 
In this model, we use as simple model as possible in order to obtain the clear view of understanding physics. 
For the full-opacity model, we found that $\dot M_{\rm TT} \sim 10^{-8} M_{\odot} {\rm yr^{-1}}, ~\dot M_{\rm FU}\sim 10^{-4} M_{\odot} {\rm yr ^{-1}}$, and $t_{\rm due}\sim 10^2 $ yr. 
These values are closer to observed ones than those for the constant opacity model. 

\begin{table}
  \caption{The property of outburst}\label{tab:comp_obs}
  \begin{center}
    \begin{tabular}{|c|c|c|c|}
    \hline
      & observations & constant opacity $\kappa =3$& full opacity table \\
     \hline
     $\dot M_{\rm TT}\rm [M_{\odot}~yr^{-1}]$ & $10^{-8}-10^{-7}$ & $10^{-12} $ & $10^{-8}$\\
     \hline
     $\dot M_{\rm FU}\rm [M_{\odot}~yr^{-1}]$ & $10^{-5}-10^{-4}$ & $10^{-5}$ & $10^{-4}$\\
      \hline
     $t_{\rm dur}$[yr] & $10^2-10^3$ & $10^3$ & $10^2$\\
          \hline
     $t_{\rm int}$[yr] & $10^3-10^5$ & $10^4-10^6$&\\
          \hline
    \end{tabular}
  \end{center}
\end{table}

\subsection{Ignored processes and future direction of this study}

First, we have simplified the condition whether MRI is active or not. 
In this paper, the condition for MRI activation is represented as $T\ge T_{\rm crit}$
since we consider that collisional ionization is effective above $T_{\rm crit}$ 
owing to the dust sublimation. 
Actually, however, the ionization rate determines activity of the MRI. 
Ionization rate has strong gradient in vertical direction 
because it is related to the density, temperature, and external ionizing sources \citep{fuj11,lan13}. 
This fact implies that $\alpha_{\rm MRI}$ strongly depends on vertical coordinate $z$. 
Some previous studies include this effect approximately by using the layered accretion model (e.g. \cite{Armitage01,Zhu09}). 
However, we use more simple treatment in which $\alpha_{\rm MRI}$ is assumed to be constant for $z$ 
because the layered accretion is not essential component for the outbursts by GML model. 
We should include the vertical structure of the disks in order to treat the effects of MRI more precisely. 
This treatment requires at least two-dimensional calculation.  

Second, the efficiency of the angular momentum transport by GI is controversial. 
There are some formula of $\alpha_{\rm GI}$ based on analytic consideration \citep{LP90} 
or fitting to numerical results \citep{kra08}. 
However, \citet{bal99} claimed that gravitational torque cannot be expressed in alpha prescription. 
Alpha prescription is founded on the assumption that the torque acts as local stress. 
Since gravity is long-range force, the assumption is not well satisfied. 
In order to establish the GML model as the robust mechanism of the outbursts, 
it is necessary to check whether the gravitational torque could be represented by the alpha prescription or not.  
To investigate this problem, non-axial symmetric calculations are necessary. 

Finally, we fix the outer radius $r_{\rm out}=5~\rm AU$ that is the sink radius of the \citet{vor13}. 
According to their simulation, 
the clumps made around $100$ AU are able to arrive at the sink radius. 
They are expected to be destroyed by the tidal force from the protostar 
if they approach closer to it than tidal radius $r_{\rm tid}$ (Nayakshin \yearcite{Nayakshin}, Tsukamoto et al. \yearcite{Tsukamoto}). 
In this paper, we adopt one-dimensional model with axial symmetry to describe the inner disk. 
In the situation that clumps exist, axial symmetry is not valid, and our model should not be used. 
Once clumps are destroyed, it is expected that their remnants spread around the protostar 
to form a disk that is moderately axial-symmetric. 
Our model is likely to be available in $r \lesssim r_{\rm tid}$, 
and thus, we should set $r_{\rm out} = r_{\rm tid}$. 
If tidal radius $r_{\rm tid}$ is smaller than the radius of protostar, 
GI model could be feasible to explain outburst because the clump is expected to fall onto the protostar directly. 
The reasonable estimation of the tidal radius for each clump is important to determine the mechanism that triggers the outbursts. 
This problem also needs non-axial symmetric calculations. 
To improve all physical processes we discussed above, 
we need three-dimensional calculations in principle. 

\section{Summary}
In this paper, we have investigated the role of the inner disk $r\lesssim 5~{\rm AU}$ from the central star in the time dependent mass accretion flow. 
In order to understand property of time variation of mass accretion rate onto the central star, we considered viscous evolution of the inner disk taking into account both of gravo-magneto limit cycle (GML) and time variable mass accretion rate onto the inner disk. 
We assumed the $\alpha$-descreption in order to treat transport of angular momentum driven by both of GI and MRI. 
Our results and findings are summarized as follows:

1. Mass accretion rate onto the protostar tends to be different from that onto the inner disk partly due to the viscous diffusion. 
In the case with temporally constant viscous parameter $\nu$, when $t_{\rm osci}<t_{\rm diff}$ mass accretion rate $\dot{M}(r)$ approaches the time-averaged value of $\dot{M}_{\rm out}$ in a single period $<\dot{M}_{\rm out}>_{\rm osci}$ as $r\rightarrow 0$, and when $t_{\rm diff}<t_{\rm osci}$ time variability of mass accretion rate remains.  

2. Outburst driven by the GML can occur under the condition that constant mass accretion rate onto the inner disk is $10^{-10}<\dot{M}_{\rm out}< 3\times 10^{-6}~{\rm M}_{\odot}~{\rm yr}^{-1}$. 
The low boundary for the steady accretion rate sensitively depends on the $\alpha_{\rm M,off}$, which we know little about. 
In this range of $\dot{M}_{\rm out}$, among the features of outbursts such as $t_{\rm dur}$, $t_{\rm int}$, $\dot{M}_{\rm FU}$, and $\dot{M}_{\rm TT}$, only the interval between outburst $t_{\rm int}$ is a function of $\dot{M}_{\rm out}$. 
Difference in $\dot{M}_{\rm out}$ does not affect $\dot{M}_{\rm FU}$ but it does affects $t_{\rm int}$. 
Large $\dot{M}_{\rm out}$ results in short $t_{\rm int}$. 
Large $\dot{M}_{\rm out}$ do not appear directly in the amplitude of $\dot{M}_{\rm in}$. 

3. Even with a fluctuated mass accretion rate onto the inner disk at $r_{\rm out}$, the GML can drive outbursts although fluctuations of $\dot{M}$ may decay when passing the inner disk inwards. 
We newly identified two modes of outburst which are spontaneous one and stimulated one. 
In the case with $<\dot{M}_{\rm out}>$ smaller than $3\times 10^{-6}~{\rm M}_{\odot} {\rm yr}^{-1}$, outburst is a spontaneous one in which $\dot{M}_{\rm in}>\dot{M}_{\rm out}$, and 
in the case with the $<\dot{M}_{\rm out}>$ greater than $3\times 10^{-6}~{\rm M}_{\odot} {\rm yr}^{-1}$, outburst is a stimulated one in which $\dot{M}_{\rm in}\sim\dot{M}_{\rm out}$, 
where $<\dot{M}_{\rm out}>$ is the averaged value of $\dot{M}_{\rm out}$ during several kyr.   

The important results in the present paper are that mass accretion rate onto the sink cell does not always appear directly in mass accretion rate onto the star (the latter determinng the accretion luminosity), but we suggest that the interval between outbursts is possibly used as a probe for mass accretion rate. 

Although we used many simplified treatment in the model, we believe our results help understand the role for the inner disk in mass accretion rate onto the star in the early phase of star formation. 












\bigskip

We are thankful to the anonymous referee for useful comments and suggestions that helped to improve the paper. 
We acknowledge to Kentaro Nagamine and Fumio Takahara for useful discussions and continuous encouragement. 
This work is partly supported by Grant-in-Aid for JSPS Fellow No. 251784 (S.S.K.). 
E. I. Vorobyov acknowledges support from the Russian Fund for Fundamental Research grant 14-02-00719.  

\appendix




\end{document}